\documentclass[aps,prb,superscriptaddress,sd,amsmath,reprint,longbibliography,amssymb]{revtex4-1}
\usepackage[utf8]{inputenc}
\usepackage[T1]{fontenc}
\usepackage{textcomp}
\usepackage{amstext}
\usepackage{graphicx}
\usepackage{subscript}
\usepackage{hyperref}
\usepackage{natbib}
\usepackage[english]{babel}
\usepackage{tablefootnote}
\usepackage[usenames]{color}

\makeatletter

\newcommand{\lyxmathsym}[1]{\ifmmode\begingroup\def\b@ld{bold}
  \text{\ifx\math@version\b@ld\bfseries\fi#1}\endgroup\else#1\fi}

\usepackage{float}
\usepackage{times}
\usepackage{subfigure}
\usepackage{color}

\makeatother

\begin{document}

\title{Growth and characterization of ultrathin cobalt ferrite films on Pt(111)}














\author{G.D. Soria}
\affiliation{Instituto de Química Física ``Rocasolano'', CSIC, Madrid E-28006,
Spain}

\author{K. Freindl}
\affiliation{Jerzy Haber Institute of Catalysis and Surface Chemistry, Polish
Academy of Sciences, 30-239 Kraków, Poland}

\author{J. E. Prieto}
\affiliation{Instituto de Química Física ``Rocasolano'', CSIC, Madrid E-28006,
Spain}

\author{A. Quesada}
\address{Instituto de Cerámica y Vidrio, CSIC, Madrid E-28049, Spain}

\author{J. de la Figuera}
\affiliation{Instituto de Química Física ``Rocasolano'', CSIC, Madrid E-28006,
Spain}

\author{N. Spiridis}
\affiliation{Jerzy Haber Institute of Catalysis and Surface Chemistry, Polish
Academy of Sciences, 30-239 Kraków, Poland}

\author{J. Korecki}
\affiliation{Jerzy Haber Institute of Catalysis and Surface Chemistry, Polish
Academy of Sciences, 30-239 Kraków, Poland}
\affiliation{AGH University of Science and Technology, Faculty of Physics and
Applied Computer Science, 30-259 Kraków, Poland}

\author{J. F. Marco}
\affiliation{Instituto de Química Física ``Rocasolano'', CSIC, Madrid E-28006,
Spain}

\begin{abstract}
CoFe$_{2}$O$_{4}$ thin films (5 nm and 20 nm thick) were grown by oxygen assisted molecular beam epitaxy on Pt(111) at 523~K and subsequently annealed at 773 K in vacuum or oxygen. They were characterized \emph{in-situ} using Auger Electron Spectroscopy, Low-Energy Electron Diffraction, Scanning Tunneling Microscopy and Conversion Electron Mössbauer Spectroscopy. The as-grown films were composed of small, nanometric grains. Annealing of the films produced an increase in the grain size and gave rise to magnetic order at room temperature, although with a fraction of the films remaining in the paramagnetic state. Annealing also induced cobalt segregation to the surface of the thicker films. The measured Mössbauer spectra at low temperature were indicative of cobalt ferrite, the both films showing very similar hyperfine patterns. Annealing in oxygen or vacuum affected the cationic distribution, which was closer to that expected for an inverse spinel in the case of annealing in an oxygen atmosphere.
\end{abstract}
\maketitle

\section{Introduction}

Hard magnetic ferrites have many technological applications, including permanent magnets, recording media, spintronic and microwave devices \cite{BrabersHandBook1995,murdock_roadmap_1992,sugimoto_past_2004,Bibes2007,HarrisIEEEmag2012,Coll2019}. Since for many of these applications the use of ferrite thin films with well-defined properties is of paramount importance, we have focused in this work on the preparation and characterization of cobalt ferrite (CoFe$_{2}$O$_{4}$, CFO) thin films. Cobalt ferrite has the highest magnetocrystalline anisotropy among the cubic ferrites, shows large magnetostriction effects and presents both a high Curie temperature and a large saturation magnetization \cite{BrabersHandBook1995}. In thin film form it has been proposed as a spin filter in spintronic devices \cite{CareyAPL2002,ChenPRB2007,ramos_influence_2007,RamosPRB2008,MichaelJPd2010,Bibes2011,TakahashiAPL2010,chen_nanoscale_2015}.

\medskip

An ideal CFO has an inverse spinel crystal structure (AB$_{2}$O$_{4}$) with space group Fd$\bar{3}$m, in which all cobalt cations (Co$^{2+}$) together with half of the iron cations (Fe$_{B}^{3+}$) occupy octahedral sites (B), while the other half of iron cations (Fe$_{A}^{3+}$) are located at tetrahedral sites (A), resulting in the chemical formula Fe$_{A}^{3+}$Fe$_{B}^{3+}$Co$_{B}^{2+}$O$_{4}$. Real samples always present a partial Co$^{2+}$ occupation of the tetrahedral sites, whose particular value depends on the preparation method and the sample history \cite{TakahashiJAP1972,MartensJPhysChemSol1985,deGraveHI1994,deGrave2013,JuanCCA2015,deSantisACB2019}. The high magnetocrystalline anisotropy is attributed to the high orbital moment of the Co$^{2+}$ cations. The net magnetization is smaller than that of the isostructural magnetite (pure iron inverse spinel), as the Co$^{2+}$ cations have a smaller spin moment compared with the Fe$^{2+}$ of the latter. A technique often used to study the cationic distribution in CFO is Mössbauer spectroscopy \cite{TakahashiJAP1972}. In this case, iron in octahedral and tetrahedral positions give rise to two sextets in the Mössbauer spectrum. However, the two sextets overlap strongly at room temperature, and low temperature experiments are required to separate both signals. Even at low temperature, the overlap is significant \cite{sanchez-arenillas_bulk_2019} requiring the application of external magnetic fields for a non-ambiguous determination of the octahedral and tetrahedral populations and, therefore, of the inversion degree \cite{deGraveHI1994,deGrave2013}. Nevertheless, Mössbauer spectroscopy remains one of the most useful techniques to characterize ferrites and, in particular, CFO.

\medskip

CFO films have been grown by many different methods, such as sol-gel process  \cite{dos_s_duque_magnetic_2001,lee_magnetic_1998}, dual ion beam sputtering  \cite{okuno_preferred_1992,prieto_epitaxial_2018}, pulsed laser deposition  \cite{MichaelAFM2012,bilovol_study_2014,ManuelPRB2016,eskandari_magnetization_2017,oujja_effect_2018}, magnetron sputtering  \cite{MichaelAFM2012,lee_surface_2003,RigatoPRB2009},  and molecular beam epitaxy to cite a few. The latter method has been performed depositing Co and Fe in atomic oxygen \cite{ramos_influence_2007,horng_magnetic_2004} or molecular oxygen \cite{martin-garcia_atomically_2015,SandraJCP2020} enviroments, followed by subsequent oxidation steps of the deposited metal layers \cite{deSantisACB2019}, by depositing Co on magnetite \cite{RaquelJCP2016} or even by annealing oxide layers \cite{RodewaldPRB2019,ThienJPCC2020}. In nearly all those cases, the substrates have been oxides or insulators. However, there are cases where the growth of thin films of CFO on metal substrates is desirable. With an applied goal in spintronic applications, it would allow studying the possible modification of magnetic domains by current flow, either due to spin-orbit or spin-transfer torque. From the characterization point of view, it allows using the full range of electron-based surface probes on ultrathin films to determine the structural, chemical and magnetic properties of the grown films. While the growth of magnetite films, i.e. pure-iron spinel with mixed valence, has been performed on a variety of metal substrates such as Au, Ag, Pt, Ru and W \cite{RankeReview,GarethReview2016}, their use has been much more scarce in the case of cobalt ferrite. Santis et al. \cite{deSantisACB2019} used a post annealing step after depositing Co and Fe on Ag(100), obtaining (100)-oriented films. Recently, we have grown cobalt ferrite films on hexagonal single crystal substrates, such as Ru(0001), by oxygen-assisted high temperature molecular beam epitaxy  \cite{martin-garcia_atomically_2015, SandraJCP2020} obtaining (111) oriented crystallites. However, in a similar manner to other spinels such as magnetite \cite{SantosJPc2009,MontiPRB2012,SandraNano2018} and nickel ferrite \cite{AnnaSciRep2018}, the growth typically does not provide single-phase films with a flat morphology. Instead, it gives rise to crystallites of non-stoichiometric CFO islands sitting on a wetting layer of Fe(II)-Co(II) mixed oxide. In many applications, a continuous uniform film is desired. Thus, in the present work, we investigate a method to obtain continuous CFO films on top of a metal substrate, Pt (111), by oxygen-assisted molecular beam epitaxy, but keeping the substrate at much lower temperatures to prevent the phase separation observed at high temperatures \cite{SandraJCP2020}. In this study, we use {\it in-situ} Mössbauer spectroscopy to identify the environment of the iron cations, and we complement it with Auger electron spectroscopy (AES), low-energy electron diffraction (LEED), and scanning tunnelling microscopy (STM).

\section{Experimental Methods}

Cobalt ferrite films were prepared and characterized {\it in-situ} in an ultra-high vacuum (UHV) multichamber system. The system included a preparation chamber equipped with a molecular beam epitaxy doser, a LEED difractometer (OCI Vacuum Microengineering), which can be also used to record AES data using it as a Retarding Field Analyser (RFA), a chamber with STM (Burleigh Instruments microscope with Digital Instruments control electronics), and a chamber devoted to Conversion Electron Mössbauer Spectroscopy (CEMS). The home-made CEMS spectrometer, fitted with a channeltron to detect the electrons emitted from the sample, and a 100 mCi Mössbauer $^{57}$Co(Rh) $\gamma$-ray source located outside the vacuum chamber, has been described in Ref.~\cite{SpiridisJPC2019}.  The spectra were measured with the incoming $\gamma$-rays along the surface normal.

The substrate was a Pt(111) single crystal. Its temperature was controlled by a PtRh-Pt thermocouple pressed against the back side of the crystal. It was cleaned by cycles of sputtering with argon ion bombardment (500 eV, 5${\mu}$A, for 35 min), flashing in UHV and annealing in atmosphere of oxygen (1$\times 10^{-7}$ mbar, 10 min, 825~K). The cycles were repeated until a sharp (1$\times$1) Pt${(111)}$ LEED pattern was obtained. Nonetheless, some residual iron contamination could be detected by CEMS after numerous preparation cycles due to iron impurities diluted deeper into the substrate. 

The cobalt ferrite films were grown by co-deposition of iron, enriched to 95$\%$ in $^{57}$Fe, and cobalt in an oxygen atmosphere on a heated Pt substrate. The oxygen partial pressure was $8\times10^{-6}$ mbar, and the substrate temperature was kept at 523~K. The growth rates of iron and cobalt and the annealing treatments for each film are explained in their respective sections.


At each preparation step, the films were characterized by LEED and AES. At selected steps, STM and CEMS measurements were performed. The latter were acquired both at room temperature (RT) and low temperature (LT, 115--125~K). The spectra and the magnetic field distributions were fitted using the Recoil program and the NORMOS code \cite{NORMOS}. As mentioned above, the Auger spectra were acquired by means of the four-grid LEED spectrometer (in RFA mode) using a primary electron beam (I$_P$) of 1.7~keV.

\section{Results}
\subsection{20 nm film}

This sample was deposited using growth rates of approximately 0.28 nm/min for Fe and 0.14~nm/min for Co, respectively, i.e. keeping the nominal Fe flux approximately twice as large as that of Co, in order to obtain a nominal film of composition close to CoFe$_{2}$O$_{4}$. Subsequently, the film was annealed at 673~K and 773~K in UHV for 15 min in both treatments.

\medskip

\begin{figure}[htb!]
\centerline{\includegraphics[width=0.50\textwidth]{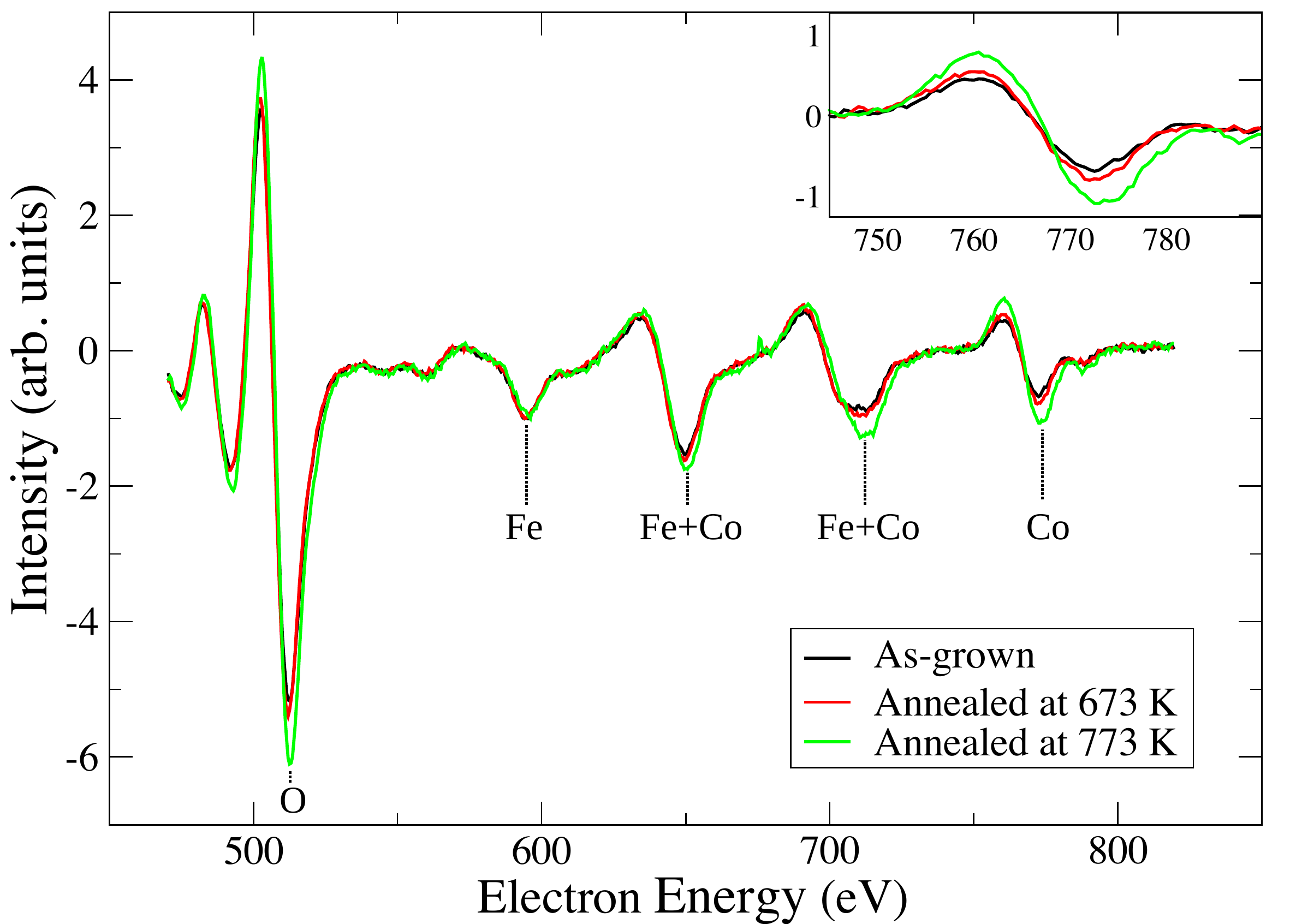}} 
\caption{Auger spectra recorded from the 20 nm film. The spectra were normalized to the intensity of the Fe peak at 598 eV. Inset: Auger Co LMM peak at 775 eV.}
\label{Auger20}
\end{figure}

Figure \ref{Auger20} shows the AES spectra recorded from the as-grown film, as well as for the film subjected to annealing at 673 and 773~K. All spectra show the expected oxygen KLL, iron LMM and cobalt LMM lines. Due to the overlap of several of the Co and Fe LMM lines, we used the peaks at 598 and 775~eV of Fe and Co, respectively, to monitor changes in the film composition. The data show an increase in the intensity of the Co LMM lines with respect to the Fe LMM lines upon annealing, going from a ratio of Fe$_{(598 eV)}$/Co$_{(775 eV)}$~0.95 to 0.81 (673~K) and 0.60 (773~K). Assuming the nominal composition for the as-prepared film, this intensity evolution implies the change of the Fe to Co ratio from approximately 2:1 to 4:3. We note that the electron beam energy used to acquire the AES spectra is rather low, 1.7~keV. The inelastic mean free path for Auger electrons at I$_P$ of 600 eV is close to 1 nm, so here we are collecting electrons coming from a surface layer 2-3 nm thick \cite{powell_practical_2020}. Therefore, the annealing treatment causes an increase in the ratio of cobalt atoms to iron atoms near the surface. This increase in cobalt atoms or decrease in iron atoms on the surface with the annealing step is also reflected in the oxygen peak (512 eV). The ratios between the intensity of the O KLL lines with respect to the Fe LMM intensity line, Fe$_{(598 eV)}$/O$_{(512 eV)}$, after growth and after the annealing treatments at 673~K and 773~K are 0.115, 0.110, and 0.095, respectively. This data trend from the Fe/O ratio confirms the decrease in the number of iron atoms on the surface upon annealing.

\begin{figure}[htb!]
\centerline{\includegraphics[width=0.5\textwidth]{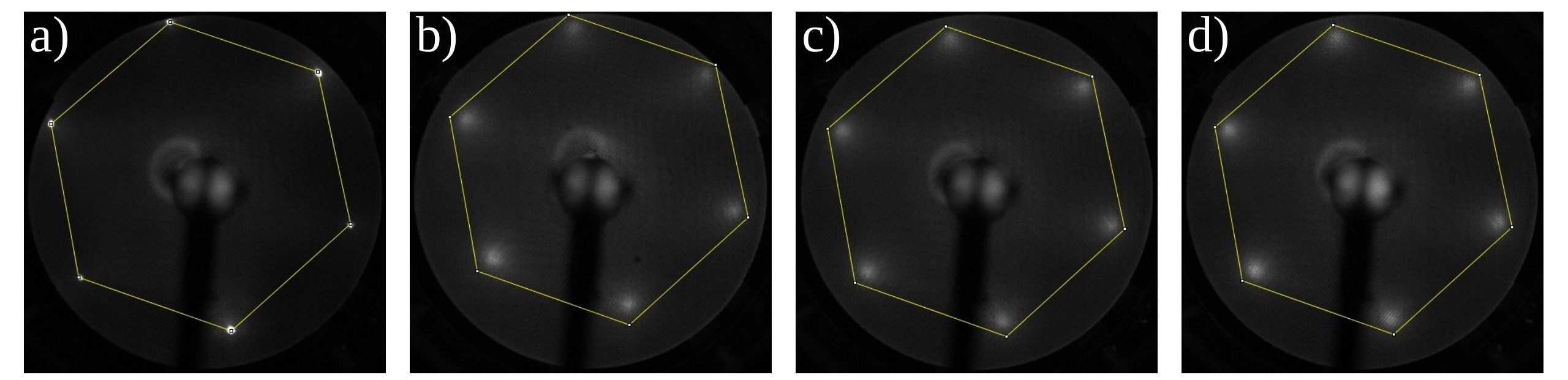}}
\caption{Diffraction patterns for: a) Platinum substrate, 20 nm CFO film, b) as-grown, c) annealed at 673 K, and d) at 773 K. LEED patterns were acquired at an energy of 66 eV.}
\label{LEED20}
\end{figure}

The LEED patterns of the 20~nm thick CFO sample recorded at different stages of its preparation are shown in figure~\ref{LEED20}. In the first panel, the pattern from the substrate is presented, showing the Pt(111) first-order spots. The as-grown film also shows a $1\times1$ pattern, with the same symmetry as the substrate and somewhat smaller spacing, which corresponds to a larger surface unit cell of 0.30~nm compared to 0.277 nm for Pt(111). No significant changes are observed in the pattern upon annealing.  Although hexagonal symmetry is expected for the (111) surface of the cobalt ferrite, the observed surface unit cell is smaller by a factor of two than that expected for the (111) plane of bulk cobalt ferrite which has a 0.59 nm periodicity. We will comment on this observation in the discussion section.    

\begin{figure}[htb]
\centerline{\includegraphics[width=0.5\textwidth]{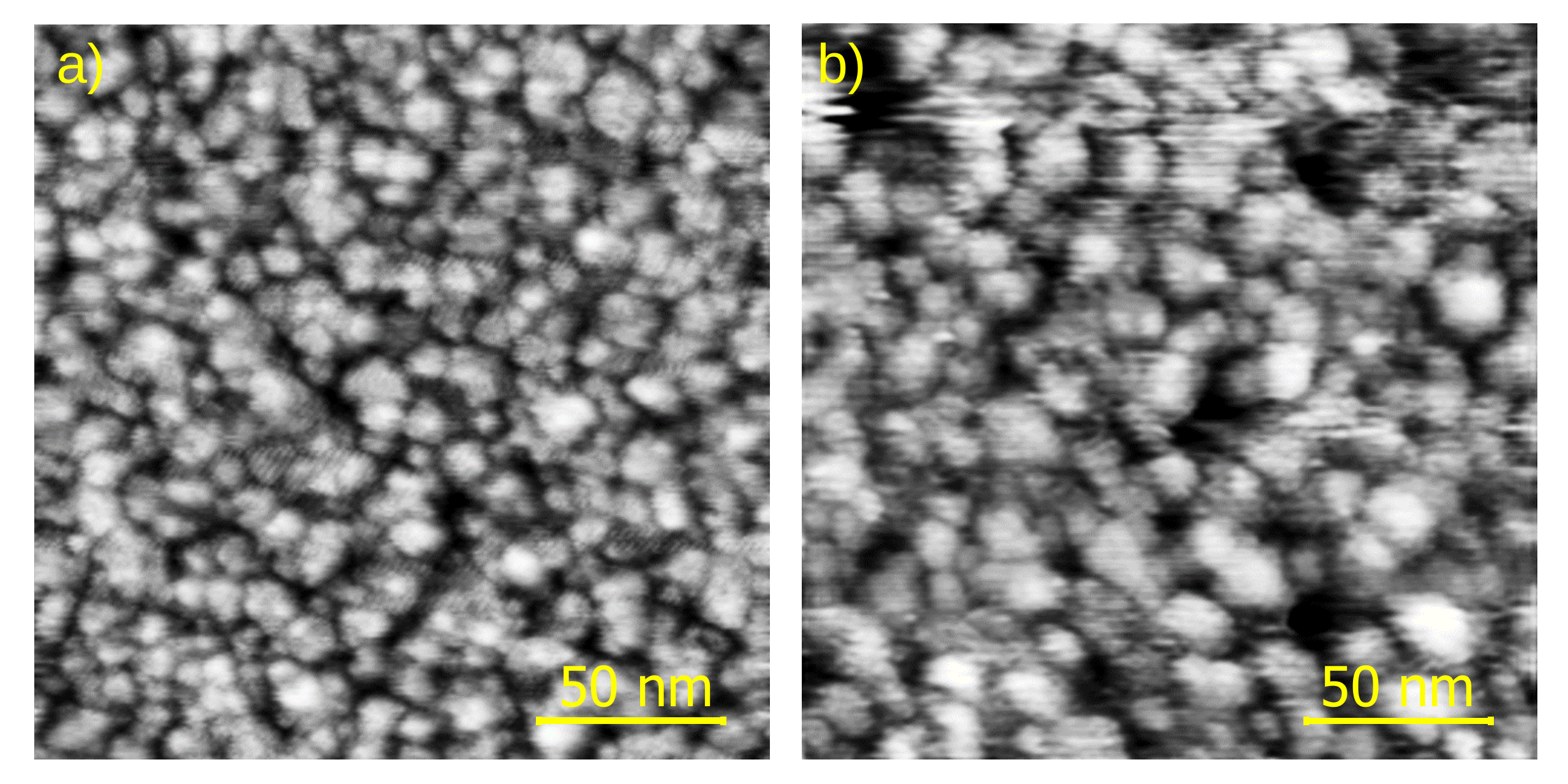}} 
\caption{STM images of the 20 nm CFO thin film: a) as-grown and b) annealed to 773 K.}
\label{STMB}
\end{figure}

Figure~\ref{STMB}a shows an STM image of the as-grown film. The image shows the presence of particles with a size around 15--20 nm and a surface rms roughness of 0.3~nm. Figure~\ref{STMB}b, which corresponds to the sample after annealing at 773~K in UHV, shows a film with somewhat larger clusters 20--25 nm in size and slightly larger 0.4~nm rms roughness.

\medskip

\begin{figure}[htb!]
\centerline{\includegraphics[width=0.5\textwidth]{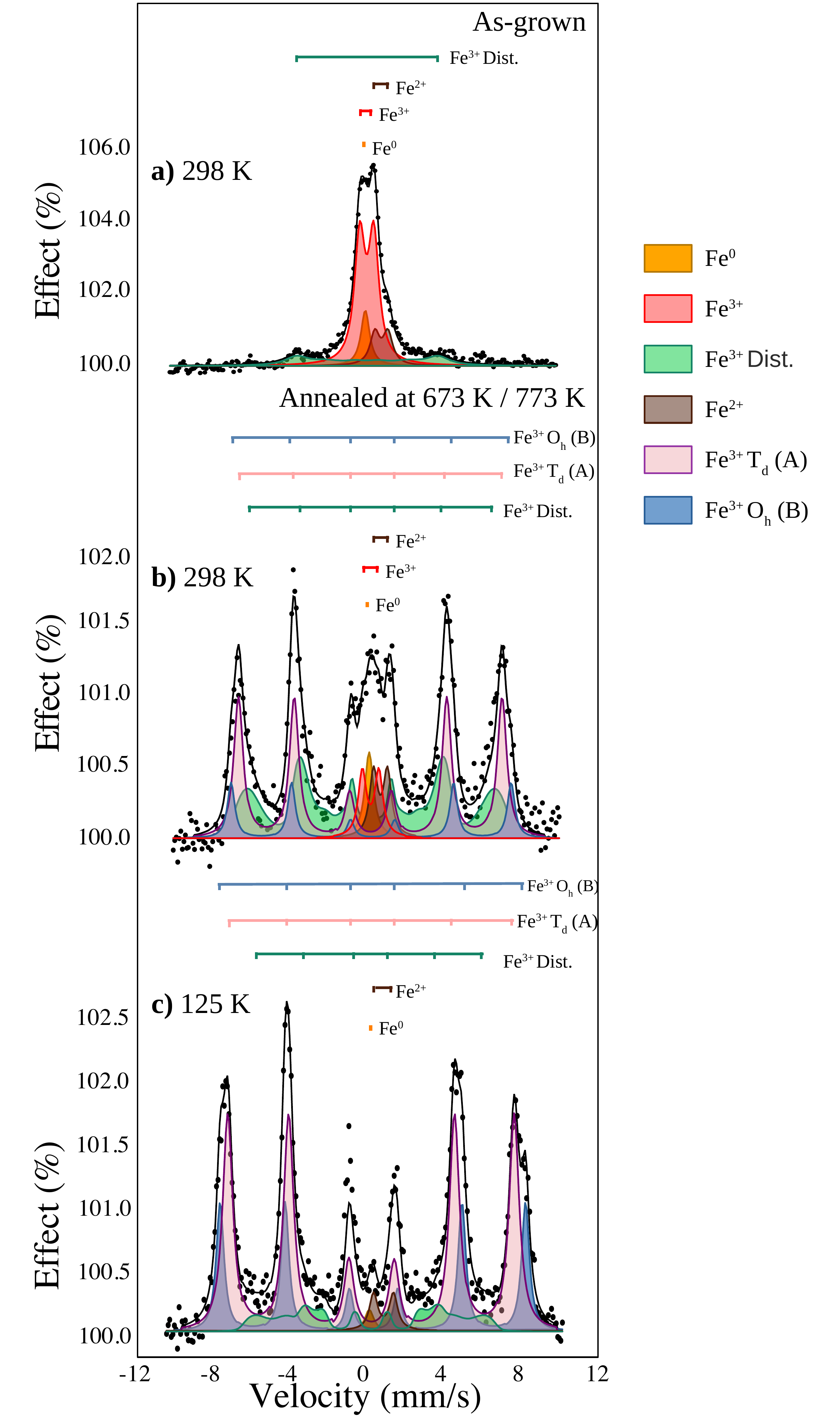}}
\caption{Mössbauer spectra obtained for the 20 nm cobalt ferrite thin film for the different stages: a) as-grown (measured at 298 K), b) and c) annealed to 673 K and 773 K, measured at 298 K and 125 K, respectively.}
\label{MossbauerB}
\end{figure}

The Mössbauer spectrum recorded from the as-grown sample is depicted on top of figure~\ref{MossbauerB}. The spectrum is dominated by intense paramagnetic signals (a singlet and two doublets) in its central part. A low intensity, broad magnetic component is also observed. The values of the hyperfine parameters for the various components obtained from the fit of this spectrum are collected in table \ref{tableMb}. The paramagnetic singlet is due to the presence of Fe$^0$ dissolved in the platinum substrate, something which is quite common in this type of systems. The most intense doublet is characteristic of a high spin Fe$^{3+}$ species in distorted octahedral oxygen coordination, while the minor doublet can be associated with a high spin Fe$^{2+}$ species also in octahedral coordination. The weak magnetic component reflects the occurrence of a broad hyperfine magnetic field distribution, most probably arising from the poor crystallinity or from the presence of a distribution of small-sized particles (as revealed by STM) in the deposited film, whose Mössbauer parameters are characteristic of Fe$^{3+}$.\\
\\
We should mention that the inclusion of a paramagnetic Fe$^{2+}$ doublet results from the evident occurrence of a bump at around +1.5 mms$^{-1}$ which makes the central “paramagnetic” part of the spectrum to be asymmetric. However, as it occurs in the case of magnetite thin films, where strong superparamagnetism is observed due to the occurrence of structural domains, this spectrum can be also fitted using hyperfine magnetic field distributions to account for superparamagnetic components. An example of this type of fit is shown in Figure 1 of the Supplementary Information. It is interesting to note (see also Table in the SI) that the isomer shifts of these magnetic field distributions are very high (0.53-0.56 mms$^{-1}$) suggesting also the presence of Fe$^{2+}$ in the film. In this case, the result would indicate that those components would have a “magnetite-like” behavior, showing the occurrence of electron hopping.  Importantly, both fitting approaches are suggestive of the presence of Fe$^{2+}$ in the as-grown film. As it happens very often, the fit of such a spectrum cannot be unique and several models are possible.  In any case, independently of the final fit chosen,  the evident relaxation components and the likely presence of Fe$^{2+}$ in the spectrum forced us to anneal the film as a means to get rid out of such contributions which are not the expected for CFO.

\medskip

Thus, the Mössbauer spectra recorded at 298 K and LT from the film annealed up to 773~K are also shown in figure~\ref{MossbauerB}. The results clearly indicate that the annealing treatment has a significant effect on the nature of the deposited film. 

The 298 K spectrum shows the presence of a broad intense magnetic component and some other small paramagnetic components (the Fe$^0$ singlet, a Fe$^{3+}$ doublet and a Fe$^{2+}$ doublet). The fit of the magnetic component can be performed in different ways. After trying different fitting models, we adopted finally a model considering two discrete magnetic sextets and a sextet with a hyperfine magnetic field distribution. The Mössbauer parameters of the discrete sextets (table \ref{tableMb}) are typical of those shown by the tetrahedral and octahedral Fe$^{3+}$ cations of oxides with spinel-related structure. The magnetic field distribution component has a difficult assignment since its isomer shift is intermediate between those of the octahedral and tetrahedral discrete sextets. Most probably, it contains both octahedral and tetrahedral contributions. The assignment of the paramagnetic components remains as explained above.

\begin{table}[htb]
\centering
\resizebox{8.5cm}{!} {
\begin{tabular}{|c|c|c|c|c|c|}
\hline 
Spectrum & Site &$\delta$ & $\Delta$ - 2$\varepsilon$ & H & Area\\
 &  & ($\pm$ 0.03 mms$^{-1}$) & ($\pm$ 0.05 mms$^{-1}$) & ($\pm$ 0.05 T) & ($\%$)\\
\hline 
As-grown & Fe$^{0}$ & 0.33 & - & - & 11 \\
298 K & Fe$^{2+}$ & 1.25 & 0.75 & - & 14 \\
 & Fe$^{3+}$ & 0.41 & 0.80 & - & 58 \\
 & Fe$^{3+}$ & 0.43 & 0.02 & 25.1 & 17\\
Annealed & Fe$^{0}$ & 0.23 & - & - & 5 \\
298 K & Fe$_{B}^{3+}$ & 0.38 & -0.09 & 49.3 & 12 \\
 & Fe$_{A}^{3+}$ & 0.30 & -0.01 & 46.4 & 37 \\
 & Fe$^{3+}$ & 0.31 & 0.93 & - & 6 \\
 & Fe$^{2+}$ & 0.87 & 0.78 & - & 6 \\
 & Fe$^{3+}$ & 0.35 & -0.04 & 44.0 (H$_{AVG}$ - 37.1)$^a$ & 34\\
Annealed & Fe$^{0}$ & 0.30 & - & - & 1 \\
125 K & Fe$_{B}^{3+}$ & 0.51 & -0.04 & 52.4 & 27 \\
 & Fe$_{A}^{3+}$ & 0.41 & -0.03 & 49.1 & 57 \\
  & Fe$^{2+}$ & 1.10 & 1.10 & - & 3 \\
 & Fe$^{3+}$ & 0.45 & -0.09 & 43.0 (H$_{AVG}$ - 34.6)$^a$ & 12\\
\hline 
\end{tabular}
}
\\
\footnotesize{$^a$ In the case of the distribution component, H corresponds to the maximum of the distribution while H$_{AVG}$ refers to the average field of distribution.}
\caption{$^{57}$Fe Mössbauer parameters obtained from the fit of the spectra shown in figure \ref{MossbauerB}. The symbols $\delta$, $\Delta$, 2$\varepsilon$, H correspond to isomer shift, quadrupole splitting, quadrupole shift and and hyperfine magnetic field, respectively. The isomer shift values are quoted relative to $\alpha$-Fe at room temperature.}
\label{tableMb}
\end{table}

The 125 K Mössbauer spectrum presents much narrower lines than the RT one and the two discrete sextets are now much better resolved. However, the spectrum still shows some relaxation character and an additional broad (characterized by a hyperfine magnetic field distribution) magnetic component, was included in the fit. Apart from this, the spectrum continues showing the small Fe$^0$ and Fe$^{2+}$ contribution.  It is well-known \cite{oujja_effect_2018,cheng_chemical_1998} that annealing promotes both cation rearrangement among the tetrahedral and octahedral sites as well as an increase in the size of the small particles which compose the film. Specifically for small particles, the Mössbauer spectra reflect the change in the magnetic behaviour by a transition from a doublet (superparamagnetic) to a sextet (magnetic character) with the temperature decrease as well as the particle size distribution causing line broadening since the magnetic moment stabilization does not occur at the same temperature for all particles.

The superparamagnetic character in thin films caused by the small size of the particles has already been reported in the literature. J.G. S. Duque \cite{dos_s_duque_magnetic_2001} studied this phenomenon for CFO films with particle sizes between 10-20 nm. Also, López et al. \cite{lopez2006magnetic} commented on the influence of the grain size (10-40 nm in average) in their CoFe$_2$O$_4$ films. In addition, Yanagihara´s group \cite{yanagihara_perpendicular_2011} working on 13 nm thick cobalt ferrite films on $\alpha$-Al$_2$O$_3$ (0001) pointed out the occurrence of a broad magnetic component which was interpreted as a result of a thermally fluctuating magnetic order near the blocking temperature associated to a superparamagnetic behavior of the films.

The values of the hyperfine parameters of the two discrete sextets are very similar to those expected for cobalt ferrite. It is worth noting that the line intensity ratio of the magnetic components is 3:3.3:1:1:3.3:3. This indicates that the film magnetization is mostly in-plane at 72 deg with respect to the $\gamma$-rays direction in average. Considering that in this particular spectrometer the average deviation from normal incidence due to the solid angle of the sample ilumination is approximately 6$^{\circ}$, the angle between the hyperfine magnetic field and the surface plane would be 18$^{\circ}$ $\pm$ 6$^{\circ}$. The relative areas of the discrete sextets (Fe$_A^{3+}$/Fe$_B^{3+}$) are in the ratio 2.1, i.e. quite far from the expected ratio (1.0) for a perfect CFO inverse spinel. This would imply that the magnetic field distribution component contains mostly octahedral contributions. We will come back to this point later.

\subsection{5 nm film}

The growth of this film was performed at rates of approximately 0.16 nm/min for Fe and 0.08~nm/min for Co, thus keeping the nominal Fe flux approximately twice as large as that of Co in order to obtain a film with nominal composition close to CoFe$_{2}$O$_{4}$. Afterwards, the film was annealed at 673~K for 15 min and subsequently at 773~K for other 15 min, with both processes being carried out in UHV. An additional annealing treatment at 773 K in oxygen ($1 \times 10 ^{-6}$ mbar) for 15 min was performed to check if this could improve the formation of a CFO inverse spinel. 

\medskip

\begin{figure}[htb!]
\centerline{\includegraphics[width=0.5\textwidth]{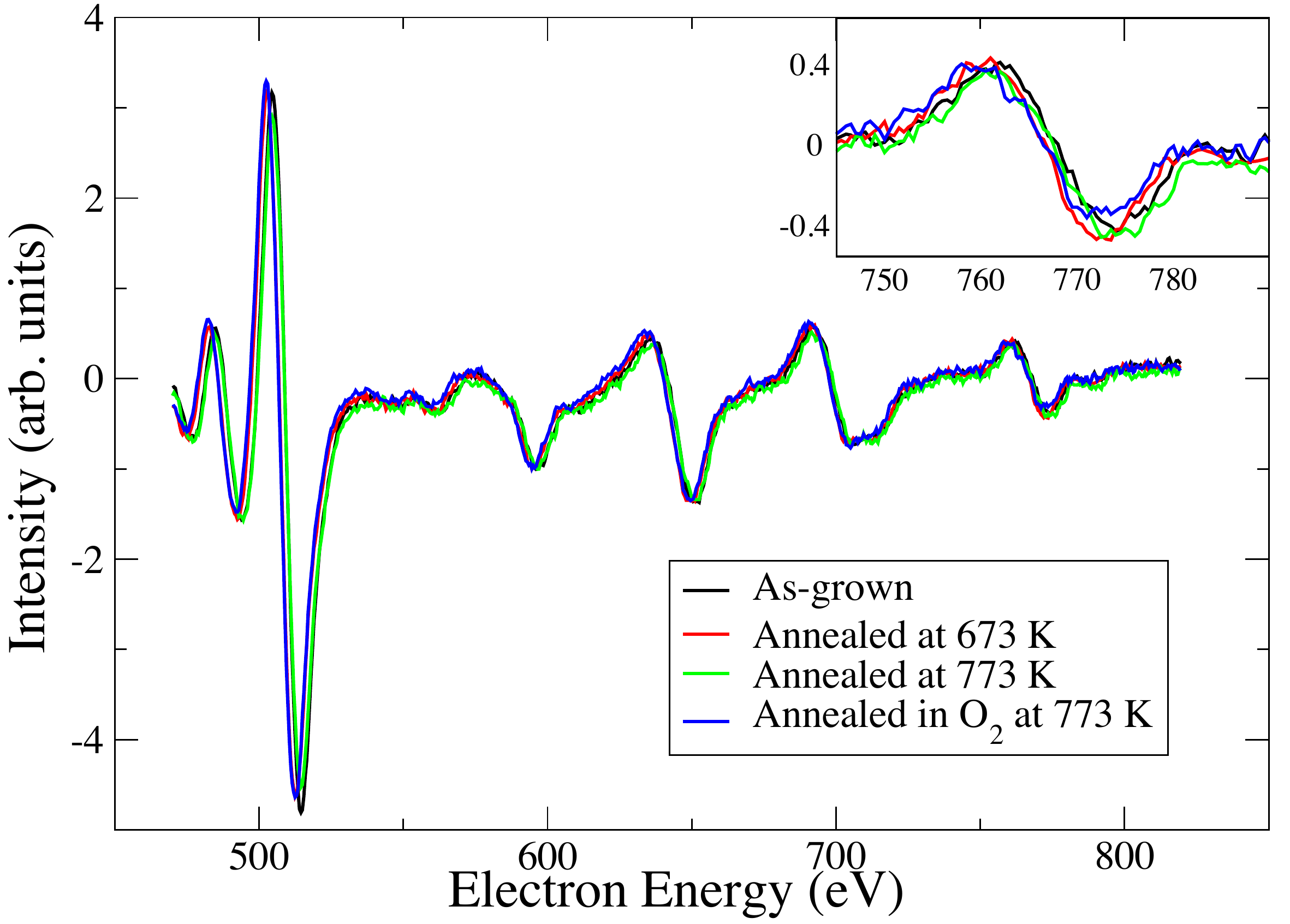}} 
\caption{Auger spectra of the 5 nm film recorded after the different treatment steps. The spectra were normalized to the intensity of the Fe peak at 598 eV. Inset: Auger Co LMM peak at 775 eV.}
\label{Auger5}
\end{figure}

The AES spectra recorded from the as-grown 5~nm thick CFO film as well as from the annealed films both in vacuum and in oxygen are shown in figure~\ref{Auger5}. Again, the lines arising from O KLL, Fe LMM and Co LMM are observed. However, unlike the case of the 20~nm thick film, which showed a clear Co enrichment upon annealing, now within the error limits of the quantitative determination, there is no clear trend.

\medskip

\begin{figure}[htb!]
\centerline{\includegraphics[width=0.5\textwidth]{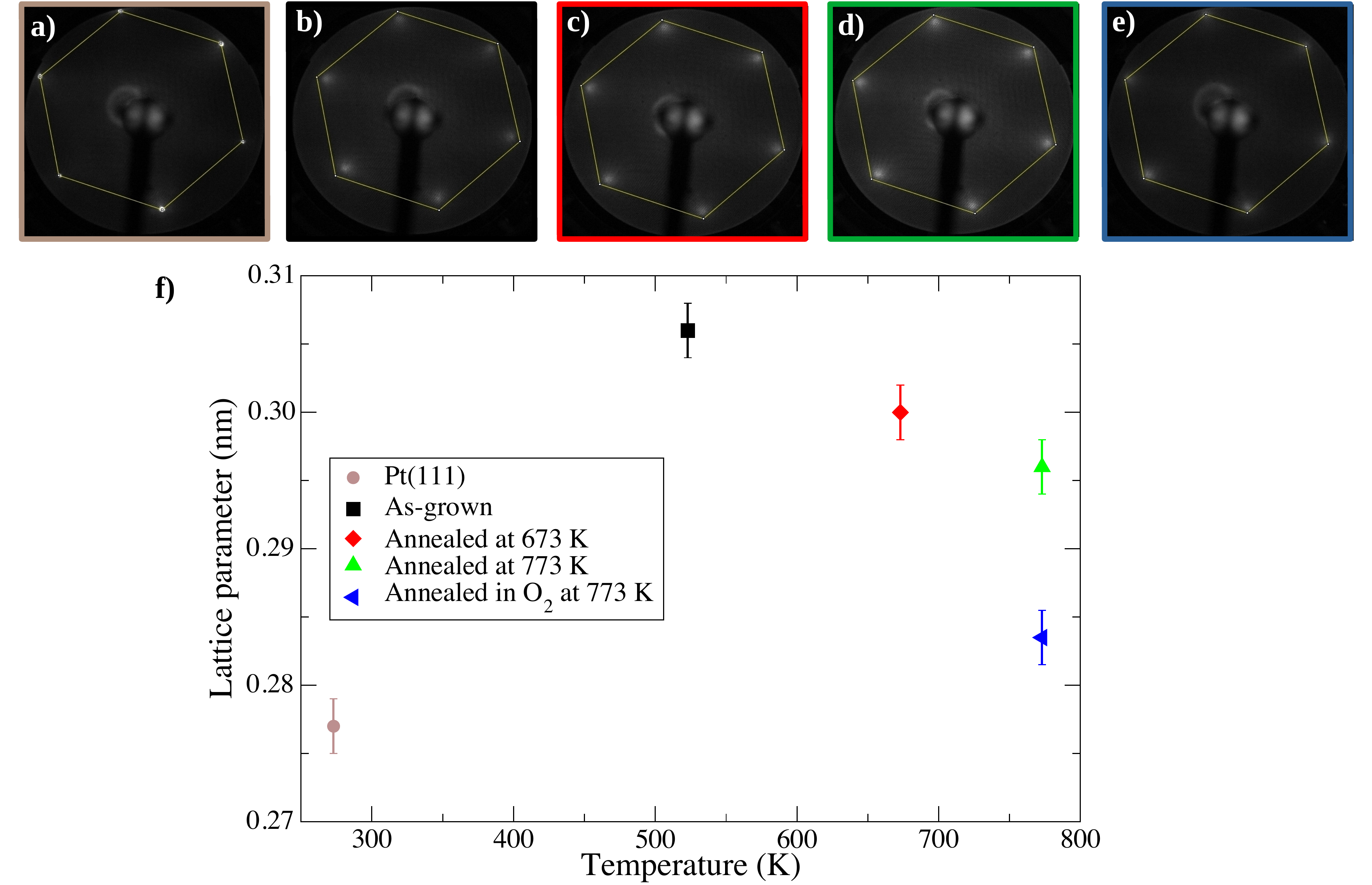}}
\caption{Top panel: LEED diffraction patterns recorded from: a) platinum substrate, 5 nm CFO film, b) as-grown, b) annealed in UHV at 673~K, d) annealed in UHV at 773~K e) annealed in O$_{2}$ at 773~K. Bottom panel: f) lattice parameters calculated from the LEED patterns after each processing step.}
\label{LEED}
\end{figure}

The LEED patterns from the as-grown and annealed films are shown in Figure~\ref{LEED}.  First, the diffraction pattern of the substrate, Pt(111), is shown (figure \ref{LEED}a). As for the thicker film, all the patterns are hexagonal with the same symmetry and orientation as the substrate. The as-grown film presents broader spots with considerably larger lattice spacing than the substrate (figure \ref{LEED}b). Annealing at 673~K and 773~K in vacuum produces a small increase in the spot separation towards the substrate lattice spacing, an effect which is more clear after annealing in O$_{2}$ (figures \ref{LEED}c, \ref{LEED}d and \ref{LEED}e, respectively). Using the Pt as a reference (0.28~nm), we show in figure~\ref{LEED}f the observed evolution of the lattice spacing in the 5 nm film.

\begin{figure}[htb!]
\centerline{\includegraphics[width=0.5\textwidth]{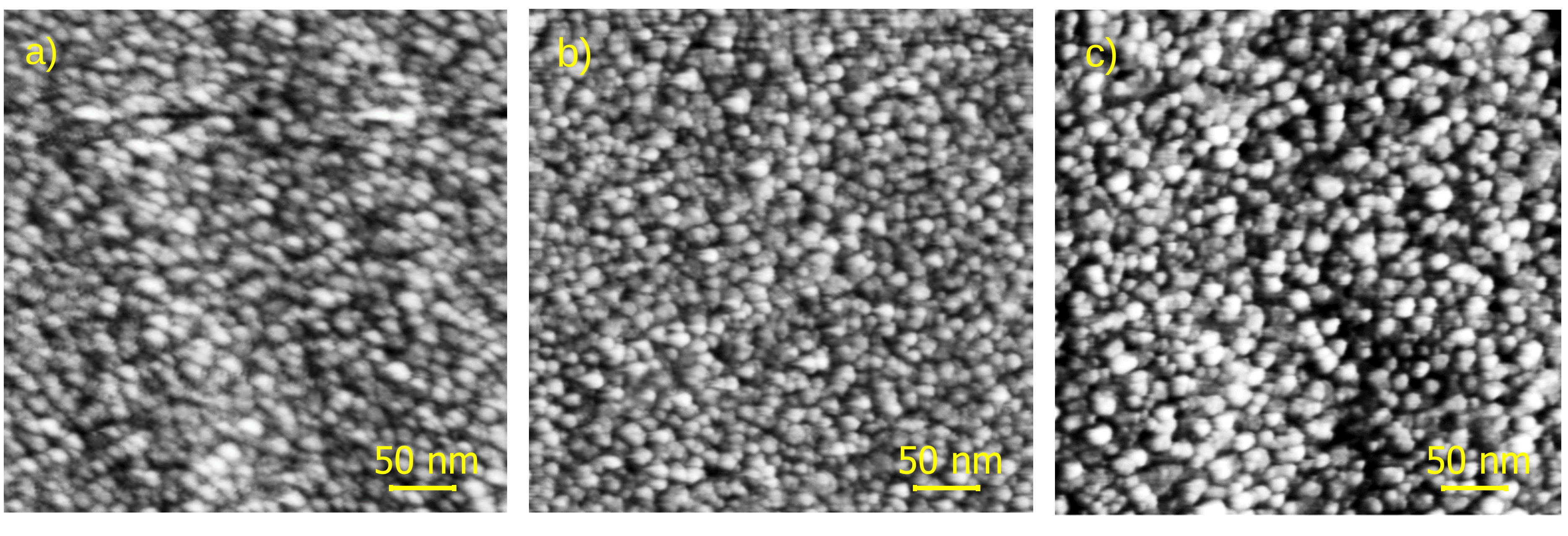}} 
\caption{STM images recorded from the 5 nm thick CFO film: a) as-grown, b) annealed at 773~K in vacuum, and c) annealed at 773~K in oxygen.}
\label{STM} 
\end{figure}

STM images recorded from the as-grown film, as well as from the film annealed at 773~K both in vacuum and in oxygen, are shown in figure \ref{STM}. They show particles 10-15 nm in size with a rms roughness of 0.3~nm, 15-18~nm and a rms roughness value of 0.4~nm and 20-25 nm with a rms roughness of 0.5~nm, respectively. 

\begin{figure}
\centerline{\includegraphics[width=0.5\textwidth]{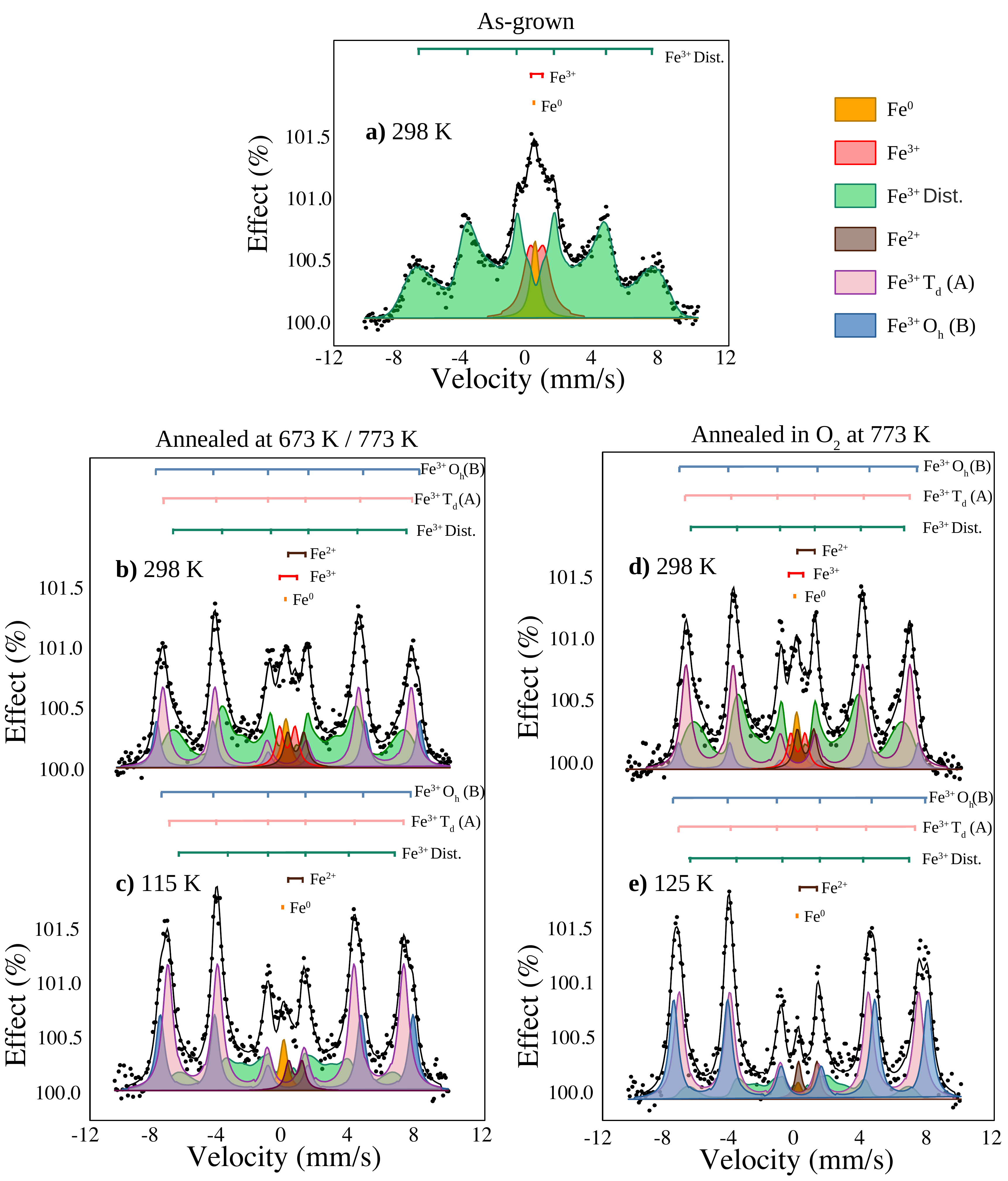}}
\caption{Mössbauer spectra recorded from the 5 nm CFO thin film for the different stages: a) as-grown film measured at 298K, b) and c) film annealed at 673 K /773 K in UHV measured at 298 K and 115 K, respectively, d) and e) film annealed in oxygen atmosphere at 773 K measured at 298 K and 125 K, respectively.}
\label{Mossbauer} 
\end{figure}

Figure \ref{Mossbauer} collects the Mössbauer data recorded from the as-grown sample at 298 K, the vacuum-annealed film both at 298 K and 115 K, and the film annealed in oxygen also both at 298 K and 125 K.

The RT spectrum recorded from the as-deposited sample is characterized by an intense, broad magnetic component that is accompanied by some other minor paramagnetic components: the Fe$^0$ singlet mentioned before and an Fe$^{3+}$ doublet in distorted octahedral oxygen coordination. Because of the broadness of the magnetic component, the presence of a Fe$^{2+}$ contribution to the spectrum has not been considered in the fit since this would imply a large uncertainty in the determination of both its Mössbauer parameters and its spectral area. The spectrum is compatible with a broad distribution of the particles sizes ranging from sufficiently small as to remain paramagnetic at RT to large enough as to give a relatively high hyperfine magnetic fields (46 T) (see Table~\ref{tableM}). 
It is worth mentioning that comparing the spectra of the as-grown 5 nm and 20 nm films, we find that the former shows only a small paramagnetic contribution. Increasing the thickness of thin films usually confer them structural and magnetic properties similar to those expected for the bulk compound. However, the opposite behavior is observed here. This result will be commented in the discussion section.


As in the case of the thicker film, the annealing treatments induce significant changes in the nature of the deposited film. The RT spectra acquired from the annealed film both in vacuum and in the presence of oxygen are very similar, and they are also similar to the RT spectrum recorded from the annealed 20 nm CFO film: they show much better defined magnetic components and significantly less intense (super)paramagnetic contributions. Therefore, they have been all fitted using a similar model and, consequently, the same considerations mentioned in the case of the 20 nm-thick annealed film are of application here. At LT, the spectra show much narrower sextet lines although, again, some magnetic relaxation is still present, hence the need of including a low-intensity magnetic component having a hyperfine magnetic field distribution. Similarly to the LT spectrum of the thick film, the 125 K spectra of these films continue to show a Fe$^{2+}$ contribution. Slight differences between the LT spectrum corresponding to the thick film and the two reported in this section refer to the Fe$_A^{3+}$/Fe$_B^{3+}$ discrete sextets intensity ratio. While in the former  this ratio amounts to 2.1, in the present case, the values obtained are 1.9 (annealed in vacuum) and 1.1 (annealed in oxygen). For the 5 nm-thick film, the line intensity ratio in the sextet is 3:3.1:1:1:3.1:3, which corresponds to an angle between the sample average magnetization and the gamma radiation direction of 70$^{\circ}$. This indicates that the sample magnetization direction forms an angle respect to the surface plane close to 20$^{\circ}$ $\pm$ 6$^{\circ}$. Therefore, in the present film the magnetization is slightly more out-of-plane than in the case of the 20 nm-thick film.

\begin{table}[htb]
\centering
\resizebox{8.5cm}{!} {
\begin{tabular}{|c|c|c|c|c|c|}
\hline 
Spectrum & Site  & $\delta$  & $\Delta$ - 2$\varepsilon$ & H & Area \\
 &  & ($\pm$ 0.03 mms$^{-1}$)  & ($\pm$ 0.05 mms$^{-1}$)  & ($\pm$ 0.05 T)  & ($\%$)  \\
\hline 
As-grown  & Fe$^{0}$ & 0.30  & -  & - & 6 \\
 & Fe$^{3+}$  & 0.40 & 0.85  & -  & 13 \\
 & Fe$^{3+}$  & 0.34  & -0.03  & 46.5 (H$_{AVG}$ - 33.7)$^a$ & 81 \\
\hline 
Annealed, RT  & Fe$^{0}$ & 0.23 & -  & -  & 4 \\
 & Fe$_{B}^{3+}$  & 0.39  & -0.05  & 49.4  & 14 \\
 & Fe$_{A}^{3+}$ & 0.29  & 0.02  & 46.7  & 31 \\
 & Fe$^{3+}$ & 0.31  & 0.93  & -  & 5 \\
 & Fe$^{2+}$ & 0.82 & 0.97 & - & 5 \\
 & Fe$^{3+}$ & 0.41 & -0.03 & 43.5 (H$_{AVG}$ - 34.9)$^a$ & 41 \\
\hline 
Annealed, 115 K & Fe$^{0}$  & 0.30 & - & -  & 3 \\
 & Fe$_{B}^{3+}$ & 0.46 & -0.08  & 51.4  & 24 \\
 & Fe$_{A}^{3+}$ & 0.36 & 0.00 & 48.0 & 47 \\
 & Fe$^{2+}$ & 0.99 & 0.88  & - & 3 \\
 & Fe$^{3+}$  & 0.41  & -0.04  & 44.0 (H$_{AVG}$ - 30.5)$^a$  & 23 \\
\hline 
Annealed in O$_{2}$, RT  & Fe$^{0}$ & 0.23 & - & - & 4 \\
 & Fe$_{B}^{3+}$  & 0.37 & -0.05  & 48.8  & 7 \\
 & Fe$_{A}^{3+}$  & 0.31  & 0.01  & 45.5  & 37 \\
 & Fe$^{3+}$  & 0.31  & 0.93  & -  & 4 \\
 & Fe$^{2+}$  & 0.86  & 1.08  & -  & 5 \\
 & Fe$^{3+}$ & 0.33 & -0.02 & 43.0 (H$_{AVG}$ - 35.5)$^a$  & 43 \\
\hline 
Annealed in O$_{2}$, 125 K & Fe$^{0}$ & 0.30 & - & - & 1 \\
& Fe$_{B}^{3+}$ & 0.49 & -0.03  & 51.9  & 37 \\
& Fe$_{A}^{3+}$ & 0.36  & -0.01  & 48.9  & 43 \\
& Fe$^{2+}$ & 0.93  & 1.21  & -  & 4 \\
& Fe$^{3+}$ & 0.40 & -0.16 & 45.2 (H$_{AVG}$ - 30.5)$^a$  & 15 \\
\hline 
\end{tabular}
}
\\
\footnotesize{$^a$ In the case of the distribution component, H corresponds to the maximum of the distribution while H$_{AVG}$ refers to the average field of distribution.}
\caption{$^{57}$Fe Mössbauer parameters obtained from the fit of the spectra
shown in figure \ref{Mossbauer}. The symbols $\delta$, $\Delta$, 2$\varepsilon$, H correspond to isomer shift, quadrupole splitting, quadrupole shift and hyperfine magnetic field, respectively. The isomer shift values are quoted relative to $\alpha$-Fe at room temperature.}
\label{tableM} 
\end{table}


\section{Discussion}

We first consider the morphology observed by STM. The as-grown films are composed of nanometric particulates having sizes of tenths of nanometers.  This explains the intense paramagnetic components and broad magnetic contributions observed in the corresponding Mössbauer spectra. However, despite having similar grain size in both as-grown films, a higher magnetic ordering is observed for the thinner film. Such evidence is surprising since one might think that an increase in the film thickness promotes a smaller (superpara)magnetic relaxation and not vice versa. This behavior might be related with the different deposition rates used to grow both films. As mentioned in the corresponding section, deposition rates for the 5 nm film were slower than for the thicker film by a factor of almost two. Taking into account the low temperature of growth (523 K), slower deposition rates would favour either the formation or larger surface aggreagates and/or induce the formation of a more ordered crystal structure.

\medskip

The LEED patterns are $1\times1$, i.e. they show the same symmetry and orientation as the substrate, with similar lattice parameter. This suggests that the films grow epitaxially on the substrate, although it is not expected that they are single-crystalline. The lattice parameter of a spinel phase is close to double in size compared to the in-plane Pt spacing (0.59~nm vs 0.27~nm). Thus one expects for an epitaxial spinel film a pattern with a periodicity close to a $2\times2$ pattern. This is in fact observed in magnetite grown on Pt(111) \cite{RankeReview,GarethReview2016,SpiridisJPC2019,SpiridisPRB2012}, or in the structure of highly perfect but non-stoichiometric cobalt ferrite grown on Ru(0001) at higher temperatures \cite{martin-garcia_atomically_2015}. There are two possible reasons for the lack of $2\times2$ diffracted beams in the LEED patterns of both as-grown and annealed films.

First, the films might be structurally quite disordered at the cationic level. Especially for the as-grown films, this is in line with the observed Mössbauer patterns. This would imply that the only unit cell that is observed is actually the one corresponding to the in-plane oxygen-oxygen spacing in the films, which is half the size of the full spinel unit cell. We note that we have reported a similar effect during the deposition of Co on Fe$_3$O$_4$(100) where at some point only the oxygen unit cell was observed in LEED \cite{RaquelJCP2016}. The second explanation is that the LEED patterns correspond to a near-surface region of the films with a rocksalt structure and thus with a Fe$_x$Co$_{1-x}$O composition. In support of this explanation, we note that the Mössbauer spectra show a small Fe$^{2+}$ contribution. There are a few reported results in support of this explanation. First, in the growth of magnetite on Pt(111) and Ru(0001), a surface reconstruction, the so-called bi-phase reconstruction, has often been observed upon molecular oxygen growth. It has been explained by a FeO-terminated spinel phase \cite{SpiridisJPC2019}. A similar reconstruction has been observed in cobalt ferrite grown at high temperature on Ru(0001) \cite{martin-garcia_atomically_2015}. Furthermore, annealing of CoO/Fe$_3$O$_4$ has also been reported to result in surface segregation of a CoO layer \cite{RodewaldPRB2019}, where it was suggested that the origin is the lower surface energy of CoO. We thus suggest that our conditions are such as to promote a rocksalt termination of the films.\\
From the two explanations proposed for the observation of a $1\times1$ LEED pattern relative to Pt, we consider more likely that the origin is cationic disorder, specially for the thinner film and the thicker one before Co segregation. Although a rock salt termination, as mentioned, has been observed as a reconstruction in (111) oriented spinels, that termination is quite thin and a moiré is observed in LEED arising from the coincidence pattern of the underlying spinel and the rocksalt structure. A similar effect should give weak $2\times2$ spots in our case arising from lower spinel layers. This is not observed. Additionally, the low growth (and annealing, for oxides) temperatures used make it likely that cationic disorder is present, something which is also supported by the Mössbauer observations.

\medskip
Annealing the samples in vacuum brings two main effects. First, it induces an enrichment in cobalt at the surface of the thicker film. This enrichment could be readily accommodated in such a Fe$_x$Co$_{1-x}$O termination, and for both, it increases the size of the particles. The latter is not too surprising, as annealing is expected to activate surface transport and promote coarsening of the films. The increase in the size of the particles is reflected in the appearance of well-developed, much better defined magnetic components in the RT Mössbauer spectra. However, the Mössbauer spectra still show the presence of some magnetic relaxation, hence the need of including a broad hyperfine magnetic field distribution, together with a Fe$^{3+}$ doublet associated with very small particles that remain in superparamagnetic state at RT.

\smallskip

It is also interesting to note that the differences among the RT Mössbauer spectra of the annealed films are really minute: they show a larger contribution of the tetrahedral sextet as compared to that of the octahedral sextet, the area corresponding to the hyperfine magnetic component amounts to ca. 34-40\% of the total spectral area and the paramagnetic contributions together do not represent more than 14\% of the spectral area. Therefore, it is difficult to conclude from the RT Mössbauer data if the difference in thickness of the films and/or the different annealing treatments (in vacuum or in the presence of oxygen atmosphere) have some influence on the structural and magnetic characteristics of the films. Fortunately, the LT Mössbauer spectra help significantly in establishing such differences.
\smallskip
It is well-known that the cation distribution of cobalt ferrite cannot be precisely determined from its RT Mössbauer spectrum. In general, the tetrahedral/octahedral site ratio is usually overestimated from such measurement. In a recent publication, we have discussed this issue in detail \cite{JuanCCA2015,sanchez-arenillas_bulk_2019}. This is mainly due to the strong overlap of the sextets corresponding to Fe$^{3+}$ in both sites, the result being very much dependent on the constraints imposed on the linewidths of both sextets during fitting. In general, the tetrahedral sextet tends to be broader, and this appears to be related to the occurrence of supertransferred magnetic fields in the spinel structure \cite{Vandenberghe1989}. Due to the supertransfer mechanism, a significant percentage of the Fe$^{3+}$ at the octahedral sites experience hyperfine magnetic fields, which can be very similar and even smaller than the average hyperfine field felt by the tetrahedral Fe$^{3+}$ cations. This broadens the tetrahedral sextet and, thus, it results in an area that is larger than that expected. This effect in the spectra is mitigated at low temperatures: due to different temperature variation of the Mössbauer parameters for both sites, the two sextets appear to be much better-resolved \cite{JuanCCA2015,sanchez-arenillas_bulk_2019}.\\

\smallskip

In the case of the spectra of the present paper, there is an additional complication since the superparamagnetic relaxation has not disappeared completely at the lowest Mössbauer acquisition temperature (125~K). The hyperfine magnetic distribution included to account for this remanent relaxation varies between 12\%-20\%  depending on the cases. Therefore, given that this broad magnetic component appears to contain both tetrahedral and octahedral contributions, it seems that it would not have a significant influence in changing the tetra/octa ratio determined at low temperature (or at least the trend observed) once it had been completely eliminated at lower temperatures (which we cannot reach with the present experimental set up). As mentioned above, these ratios are 2.1, 1.9 and 1.1 for the 20 nm-thick annealed in vacuum and 5 nm-thick film annealed in vacuum and annealed in an oxygen atmosphere, respectively. Such large values (for a perfectly inverse spinel a value of 1.0 would be expected) are usually obtained from the evaluation of low-temperature Mössbauer spectra recorded in the absence of applied magnetic field from cobalt ferrite \cite{JuanCCA2015,sanchez-arenillas_bulk_2019,prieto_epitaxial_2018}. This has been observed not only for films but also for the bulk material, and although some explanations have been advanced \cite{JuanCCA2015,sanchez-arenillas_bulk_2019} the final reason remains uncertain. In any case, from the results described here, it seems that the thickness of the film does not play a relevant role in the characteristics of the film and that, contrarily, annealing in oxygen has a significantly larger influence on the cation distribution of the final cobalt ferrite.

\medskip

The annealed films studied in this paper present significantly larger superparamagnetic relaxation than CFO films of comparable thickness prepared by IBAD \cite{JuanCCA2015} but are comparable in this respect with CFO films prepared by UV-PLD \cite{oujja_effect_2018}. This reflects once again the influence of the preparation conditions and how the properties of the films can be modified using different deposition methods, substrate temperature and composition of the environment atmosphere. It seems, however, that among all the preparation parameters, annealing at high temperatures after deposition, or deposition on a substrate maintained at high temperature, is one of the most relevant to obtain ``genuine'' or ``standard'' CFO films and that this appears to be relatively independent of the substrate. However, annealing or growing at much higher temperatures (over 1000~K) provides highly perfect structural cobalt ferrite \cite{martin-garcia_atomically_2015} but at the cost of phase separation on the film between non-stoichiometric cobalt ferrite and a rocksalt oxide \cite{SandraJCP2020}. Finer tuning of the films before reaching such a high temperature can be achieved by the proper choice of the deposition atmosphere and/or method.

These ferrite ultrathin films grown on a metallic substrate are relatively unexplored and relevant  to consider for spintronic applications. A cobalt ferrite film might provide an efficient spin-polarized current from an unpolarized current injected from the metallic substrate due to the spin-filtering effect. A possible consequence of such current flow could be to reverse the magnetization of the CFO magnetic domains, of use in magnetic information storage devices \cite{chen_nanoscale_2015,deSantisACB2019,takahashi_high_2010,ostler_ultrafast_2012}.

\section{Summary}

Cobalt ferrite thin films of different thicknesses (5 and 20 nm) have been synthesized by molecular beam epitaxy on Pt(111) and characterized {\it in-situ} under UHV conditions. Deposition at 523 K gives rise to superparamagnetic/poorly crystalline thin films composed by a distribution of Fe$^{3+}$ containing-particles of different sizes in the nanometer scale. The results show that together with the deposition of these Fe$^{3+}$-containing particles, a minor Fe$^{2+}$ contribution develops. The thinner film presents higher magnetic ordering than the thicker sample. This might be due to the lower deposition rate employed, which allows a cation distribution more similar to canonical CFO. Annealing in vacuum to 663 K and 773~K promotes an increase in size of the Fe$^{3+}$ particles which results in the development of magnetic ordering at RT, although the heating treatment has not completely eliminated the superparamagnetic relaxation. Annealing of the films also produces a cobalt enrichment of the thicker films surface, probably associated with the segregation of CoO. The low temperature Mössbauer spectra recorded from the various films indicate differences in the cation distribution of the deposited CFO films. In particular, annealing in oxygen atmosphere appears to promote the formation of a CFO film with a cation distribution close to that expected for an inverse spinel.

\section*{Acknowledgments}

This work is supported by the Spanish Ministry of Science, Innovation and Universities through Projects RTI2018-095303-B-C51, RTI2018-095303-B-C53, MAT2017-86450-C4-1-R, RTI2018-095303-A-C52, through the Ramón y Cajal Contract RYC-2017-23320 and RTI2018-095303-C52 (MCIU/AIE/FEDER,EU), by the Regional Government of Madrid through project S2018-NMT-4321 and by the PROM Programme - International scholarship exchange of PhD students and academics; and by the European Commission through the H2020 Project no. 720853 (AMPHIBIAN).

\bibliographystyle{h-physrev.bst}
\bibliography{Cobaltferrite}

\providecommand{\newblock}{}
\begin{thebibliography}{10}
\expandafter\ifx\csname url\endcsname\relax
  \def\url#1{{\tt #1}}\fi
\expandafter\ifx\csname urlprefix\endcsname\relax\def\urlprefix{URL }\fi
\providecommand{\eprint}[2][]{\url{#2}}

\bibitem{BrabersHandBook1995}
Brabers V~A~M 1995 {Progress} in spinel ferrite research {\em Handbook of
  {Magnetic} {Materials}\/} vol~8 pp 189--324

\bibitem{murdock_roadmap_1992}
Murdock E, Simmons R and Davidson R 1992 {\em IEEE Transactions on Magnetics\/}
  {\bf 28} 3078--3083

\bibitem{sugimoto_past_2004}
Sugimoto M 2004 {\em Journal of the American Ceramic Society\/} {\bf 82}
  269--280

\bibitem{Bibes2007}
Bibes M and Barthelemy A 2007 {\em Ieee Trans. Elect. Dev.\/} {\bf 54}
  1003--1023

\bibitem{HarrisIEEEmag2012}
Harris V~G 2012 {\em IEEE Transactions on Magnetics\/} {\bf 48} 1075--1104

\bibitem{Coll2019}
Coll M, Fontcuberta J, Althammer M, Bibes M, Boschker H, Calleja A and Cheng
  G~~F~M 2019 {\em App. Surf. Sci.\/} {\bf 482} 1--93

\bibitem{CareyAPL2002}
Carey M~J, Maat S, Rice P, Farrow R~F~C, Marks R~F, Kellock A, Nguyen P and
  Gurney B~A 2002 {\em Appl. Phys. Lett.\/} {\bf 81} 1044--1046

\bibitem{ChenPRB2007}
Chen Y~F and Ziese M 2007 {\em Phys. Rev. B\/} {\bf 76} 014426

\bibitem{ramos_influence_2007}
Ramos A~V, Moussy J~B, Guittet M~J, Gautier-Soyer M, Gatel C, Bayle-Guillemaud
  P, Warot-Fonrose B and Snoeck E 2007 {\em Physical Review B\/} {\bf 75}
  224421

\bibitem{RamosPRB2008}
Ramos A~V, Santos T~S, Miao G~X, Guittet M~J, Moussy J~B and Moodera J~S 2008
  {\em Phys. Rev. B\/} {\bf 78} 180402

\bibitem{MichaelJPd2010}
Foerster M, Rigato F, Bouzehouane K and Fontcuberta J 2010 {\em J. Phys. D:
  Appl. Phys.\/} {\bf 43} 295001

\bibitem{Bibes2011}
Bibes M, Villegas J~E and Barthélémy A 2011 {\em Adv. Phys.\/} {\bf 60} 5--84

\bibitem{TakahashiAPL2010}
Takahashi Y~K, Kasai S, Furubayashi T, Mitani S, Inomata K and Hono K 2010 {\em
  Appl. Phys. Lett.\/} {\bf 96} 072512

\bibitem{chen_nanoscale_2015}
Chen X, Zhu X, Xiao W, Liu G, Feng Y~P, Ding J and Li R~W 2015 {\em ACS Nano\/}
  {\bf 9} 4210--4218

\bibitem{TakahashiJAP1972}
Takahashi M and Fine M~E 1972 {\em Journal of Applied Physics\/} {\bf 43}
  4205--4216

\bibitem{MartensJPhysChemSol1985}
Martens J~W~D, Peeters W~L, van Noort H~M and Erman M 1985 {\em J. Phys. Chem.
  Sol.\/} {\bf 46} 411--416

\bibitem{deGraveHI1994}
de~Bakker P, Vandenberghe R and De~Grave E 1994 {\em Hyperfine Interactions\/}
  {\bf 94} 2023--2027

\bibitem{deGrave2013}
Le~Trong H, Presmanes L, De~Grave E, Barnabé A, Bonningue C and Tailhades P
  2013 {\em J. Magn. Magn. Mater.\/} {\bf 334} 66--73

\bibitem{JuanCCA2015}
de~la Figuera J, Quesada A, Martín-García L, Sanz M, Oujja M, Castillejo M,
  Mascaraque A, T~N’Diaye A, Foerster M, Aballe L and F~Marco J 2015 {\em
  Croatica Chemica Acta\/} {\bf 88} 453--460

\bibitem{deSantisACB2019}
De~Santis M, Bailly A, Coates I, Grenier S, Heckmann O, Hricovini K, Joly Y,
  Langlais V, Ramos A~Y, Richter C, Torrelles X, Garaudée S, Geaymond O and
  Ulrich O 2019 {\em Acta Cryst B\/} {\bf 75}

\bibitem{sanchez-arenillas_bulk_2019}
Sánchez-Arenillas M, Oujja M, Moutinho F, de~la Figuera J, Cañamares M~V,
  Quesada A, Castillejo M and Marco J~F 2019 {\em Appl. Surf. Sci.\/} {\bf 470}
  917--922

\bibitem{dos_s_duque_magnetic_2001}
dos S~Duque J~G, Macêdo M~A, Moreno N~O, Lopez J~L and Pfanes H~D 2001 {\em
  Journal of Magnetism and Magnetic Materials\/} {\bf 226-230} 1424--1425

\bibitem{lee_magnetic_1998}
Lee J~G, Park J~Y, Oh Y~J and Kim C~S 1998 {\em Journal of Applied Physics\/}
  {\bf 84} 2801--2804

\bibitem{okuno_preferred_1992}
Okuno S~N, Hashimoto S and Inomata K 1992 {\em Journal of Applied Physics\/}
  {\bf 71} 5926--5929

\bibitem{prieto_epitaxial_2018}
Prieto P, Marco J~F, Prieto J~E, Ruiz-Gomez S, Perez L, del Real R~P, Vázquez
  M and de~la Figuera J 2018 {\em Applied Surface Science\/} {\bf 436}
  1067--1074

\bibitem{MichaelAFM2012}
Foerster M, Iliev M, Dix N, Martí X, Barchuk M, Sánchez F and Fontcuberta J
  2012 {\em Adv. Funct. Mater.\/} {\bf 22} 4344--4351

\bibitem{bilovol_study_2014}
Bilovol V, Pampillo L~G and Saccone F~D 2014 {\em Thin Solid Films\/} {\bf 562}
  218--222

\bibitem{ManuelPRB2016}
Valvidares M, Dix N, Isasa M, Ollefs K, Wilhelm F, Rogalev A, Sánchez F,
  Pellegrin E, Bedoya-Pinto A, Gargiani P, Hueso L~E, Casanova F and
  Fontcuberta J 2016 {\em Phys. Rev. B\/} {\bf 93} 214415

\bibitem{eskandari_magnetization_2017}
Eskandari F, Porter S~B, Venkatesan M, Kameli P, Rode K and Coey J~M~D 2017
  {\em Physical Review Materials\/} {\bf 1} 074413

\bibitem{oujja_effect_2018}
Oujja M, Martín-García L, Rebollar E, Quesada A, García M~A, Fernández J~F,
  Marco J~F, de~la Figuera J and Castillejo M 2018 {\em Applied Surface
  Science\/} {\bf 452} 19--31

\bibitem{lee_surface_2003}
Lee J~G, Chae K~P and Sur J~C 2003 {\em Journal of Magnetism and Magnetic
  Materials\/} {\bf 267} 161--167

\bibitem{RigatoPRB2009}
Rigato F, Geshev J, Skumryev V and Fontcuberta J 2009 {\em Journal of Applied
  Physics\/} {\bf 106} 113924

\bibitem{horng_magnetic_2004}
Horng L, Chern G, Chen M~C, Kang P~C and Lee D~S 2004 {\em Journal of Magnetism
  and Magnetic Materials\/} {\bf 270} 389--396

\bibitem{martin-garcia_atomically_2015}
Martín-García L, Quesada A, Munuera C, Fernández J~F, García-Hernández M,
  Foerster M, Aballe L and de~la Figuera J 2015 {\em Advanced Materials\/} {\bf
  27} 5955--5960

\bibitem{SandraJCP2020}
Ruiz-Gómez S, Mandziak A, Prieto J~E, Aristu M, Trapero E~M, Soria G~D,
  Quesada A, Foerster M, Aballe L and de~la Figuera J 2020 {\em J. Chem.
  Phys.\/} {\bf 152} 074704

\bibitem{RaquelJCP2016}
Gargallo-Caballero R, Martín-García L, Quesada A, Granados-Miralles C,
  Foerster M, Aballe L, Bliem R, Parkinson G~S, Blaha P, Marco J~F and Figuera
  J~d~l 2016 {\em J. Chem. Phys.\/} {\bf 144} 094704

\bibitem{RodewaldPRB2019}
Rodewald J, Thien J, Pohlmann T, Hoppe M, Timmer F, Bertram F, Kuepper K and
  Wollschläger J 2019 {\em Phys. Rev. B\/} {\bf 100} 155418

\bibitem{ThienJPCC2020}
Thien J, Bahlmann J, Alexander A, Hoppe M, Pohlmann T, Ruwisch K, Meyer C,
  Bertram F, Küpper K and Wollschläger J 2020 {\em J. Phys. Chem. C\/} {\bf
  124} 23895--23904

\bibitem{RankeReview}
Weiss W and Ranke W 2002 {\em Prog. Surf. Sci.\/} {\bf 70} 1--151

\bibitem{GarethReview2016}
Parkinson G~S 2016 {\em Surf. Sci. Rep.\/} {\bf 71} 272--365

\bibitem{SantosJPc2009}
Santos B, Loginova E, Mascaraque A, Schmid A, McCarty K and de~la Figuera J
  2009 {\em J. Phys-Cond. Mat.\/} {\bf 21} 314011

\bibitem{MontiPRB2012}
Monti M, Santos B, Mascaraque A, Rodríguez de~la Fuente O, Niño M~A, Menteş
  T~O, Locatelli A, McCarty K~F, Marco J~F and de~la Figuera J 2012 {\em Phys.
  Rev. B\/} {\bf 85}

\bibitem{SandraNano2018}
Ruiz-Gomez S, Perez L, Mascaraque A, Quesad A, Prieto P, Palacio I,
  Martin-Garcia L, Foerster M, Aballe L and de~la Figuera J 2018 {\em
  Nanoscale\/} {\bf 10} 5566--5573

\bibitem{AnnaSciRep2018}
Mandziak A, Figuera J~d~l, Ruiz-Gómez S, Soria G~D, Pérez L, Prieto P,
  Quesada A, Foerster M and Aballe L 2018 {\em Sci. Rep.\/} {\bf 8} 17980

\bibitem{SpiridisJPC2019}
Spiridis N, Freindl K, Wojas J, Kwiatek N, Madej E, Wilgocka-Ślęzak D,
  Dróżdż P, Ślęzak T and Korecki J 2019 {\em J. Phys. Chem. C\/} {\bf 123}
  4204--4216

\bibitem{NORMOS}
Brand R 1995   1--123

\bibitem{powell_practical_2020}
Powell C~J 2020 {\em Journal of Vacuum Science and Technology A\/} {\bf 38}
  023209

\bibitem{cheng_chemical_1998}
Cheng F, Peng Z, Liao C, Xu Z, Gao S, Yan C, Wang D and Wang J 1998 {\em Solid
  State Communications\/} {\bf 107} 471--476

\bibitem{lopez2006magnetic}
L{\'o}pez J, Serrano W~P and Pfannes H 2006 {\em Rev. Colomb. F{\'\i}s\/} {\bf
  38} 1074--1077

\bibitem{yanagihara_perpendicular_2011}
Yanagihara H, Uwabo K, Minagawa M, Kita E and Hirota N 2011 {\em Journal of
  Applied Physics\/} {\bf 109} 07C122

\bibitem{SpiridisPRB2012}
Spiridis N, Wilgocka-Ślęzak D, Freindl K, Figarska B, Giela T, Młyńczak E,
  Strzelczyk B, Zając M and Korecki J 2012 {\em Phys. Rev. B\/} {\bf 85}
  075436

\bibitem{Vandenberghe1989}
Vandenberghe R~E and Grave E~D 1989 Mössbauer {Effect} {Studies} of {Oxidic}
  {Spinels} {\em Mössbauer {Spectroscopy} {Applied} to {Inorganic}
  {Chemistry}\/} ({\em Modern {Inorganic} {Chemistry}\/} no~3) pp 59--182

\bibitem{takahashi_high_2010}
Takahashi Y~K, Kasai S, Furubayashi T, Mitani S, Inomata K and Hono K 2010 {\em
  Applied Physics Letters\/} {\bf 96} 072512

\bibitem{ostler_ultrafast_2012}
Ostler T~A, Barker J, Evans R~F~L, Chantrell R~W, Atxitia U, Chubykalo-Fesenko
  O, El~Moussaoui S, Le~Guyader L, Mengotti E, Heyderman L~J, Nolting F,
  Tsukamoto A, Itoh A, Afanasiev D, Ivanov B~A, Kalashnikova A~M, Vahaplar K,
  Mentink J, Kirilyuk A, Rasing T and Kimel A~V 2012 {\em Nature
  Communications\/} {\bf 3} 666

\end{thebibliography}

\end{document}